\documentclass[Afour,sageh,times]{sagej}

\usepackage{moreverb,url}
\usepackage{draftwatermark}

\SetWatermarkText{DRAFT} % Change the text if needed
\SetWatermarkScale{1.0}  % Adjust the scale (1.0 is default)
\SetWatermarkColor[gray]{0.95}  % Set the color (gray, light gray is 0.9)

\usepackage[colorlinks,bookmarksopen,bookmarksnumbered,citecolor=red,urlcolor=red]{hyperref}

\newcommand\BibTeX{{\rmfamily B\kern-.05em \textsc{i\kern-.025em b}\kern-.08em
T\kern-.1667em\lower.7ex\hbox{E}\kern-.125emX}}

\newcommand{\anon}[1]{Anonymized Name}
\newcommand{\credit}{Image credit is \'Angel Acosta-Col\'on.}

\def\volumeyear{2024}

\begin{document}

\runninghead{M\'endez and Acosta-Col\'on}

\title{The Fear of Halley's Comet Visit in 1910 is Preserved in a Cave in Southern Puerto Rico}

\author{Abel M\'endez\affilnum{1,3} and \'Angel Acosta-Col\'on\affilnum{2,3}}

\affiliation{\affilnum{1}Planetary Habitability Laboratory (PHL), University of Puerto Rico at Arecibo, Arecibo, Puerto Rico, 00614\\
\affilnum{2}Karst and Cave Research and Education Group (KCREG), University of Puerto Rico at Arecibo, Arecibo, Puerto Rico, 00614 \\
\affilnum{3}Department of Physics and Chemistry, University of Puerto Rico at Arecibo, Arecibo, Puerto Rico, 00614}

\corrauth{Abel Méndez}

\email{abel.mendez@upr.edu}

\begin{abstract}
This paper explores a unique cave art found in southern Puerto Rico that depicts a comet over a tomb. Through interdisciplinary methods, including art interpretation, historical documentation, and demographic analysis, this study uncovered the artist's identity, the societal context of the period, and the potential motivations behind the creation of this art. The investigation revealed a connection to the passage of Halley's Comet in 1910 and the widespread panic it induced.
\end{abstract}

\keywords{Halley's Comet, Cave Art, Puerto Rico}

\maketitle

\section{Introduction}

Halley's Comet's visitation in 1910 stands as one of the most remarkable astronomical events of the early 20th century, captivating observers worldwide with its spectacular display. As a periodic comet visible from Earth approximately every 76 years, Halley's Comet has been documented by various civilizations over millennia, but its 1910 appearance was particularly significant due to advances in both observational technology and scientific understanding of cometary phenomena at the time \citep{legrand1986comet}.

The comet's visibility to the naked eye in April and May of 1910 provided an awe-inspiring spectacle, with its bright nucleus and long, sweeping tail visible across the night sky. This period marked a high point in communal astronomical engagement, with people around the world watching the skies to catch a glimpse of Halley's Comet \citep{turner1910halley}. The event also underscored the global nature of astronomical phenomena, bridging cultural and geographic divides through a shared celestial experience.

In the lead-up to its 1910 perihelion passage, Halley's Comet became the focus of intense public interest and scientific study. Astronomers used the latest telescopic technology to observe the approach of the comet, while the burgeoning field of spectroscopy allowed scientists to analyze the composition of the comet, revealing the presence of cyanide gas in its tail \citep{swings1965cometary}. This discovery, while a milestone in the study of cometary atmospheres, inadvertently fueled widespread public fear and fascination, as sensationalist media reports speculated on the potential effects of Earth passing through the comet's tail \citep{yeomans1991comets}.

The approach and subsequent passing of Halley's Comet near Earth on May 19, 1910, sparked significant alarm across the globe. Reports, such as one from The New York Times in February of that year, sensationalized the event with claims that the comet's tail contained deadly cyanide gases, echoing concerns raised by astronomer Camille Flammarion \citep{flammarion1910recontre}. These assertions led to widespread panic, despite reassurances from the broader astronomical community that the comet posed no real threat to life on Earth.

In a climate of fear, many individuals took extreme measures to safeguard themselves, going as far as to seal off their homes in an attempt to prevent any potential harm from the comet's gases. This episode reflects the powerful influence of scientific speculation and the media on public perception and behavior, especially when faced with the unknown aspects of celestial events.

The arrival of Halley's Comet was also highlighted in the local press in Puerto Rico. However, there is little anecdotal evidence for the local perception of the astronomical event outside that of the press. This study focuses on an illustrative response to Halley's Comet in 1910, found in a secluded cave in southern Puerto Rico.\endnote{The name of the cave and its exact location in Puerto Rico was intentionally concealed for conservation purposes. Contact Ángel Acosta-Cólon (\href{mailto:angel.acosta@upr.edu}{angel.acosta@upr.edu}) for more information.}

This study opens avenues for further investigation into the artist and the broader social and cultural networks through which knowledge of the comet's passage was disseminated and commemorated across Puerto Rico. Such considerations introduce a layer of complexity to the interpretation of the artwork, suggesting a dynamic interplay of personal, local, and historical narratives in the representation and understanding of astronomical events.

\section{Puerto Rico in the 1910s}

In 1910, Puerto Rico found itself at a pivotal juncture in its rich and complex history. A mere twelve years after the Spanish-American War of 1898, which transferred sovereignty from Spain to the United States, the island was navigating the challenges and opportunities of its new status as an unincorporated territory of the United States. This period was characterized by significant socio-political and economic transformations that would shape the island's trajectory in the 20th century.

The early 1900s marked the beginning of American influence on the island's infrastructure, education system, and economic policies. The Foraker Act of 1900 and the Jones-Shafroth Act of 1917, which granted Puerto Ricans US citizenship, were critical in defining the political and civil rights of the island's inhabitants. Despite these changes, Puerto Rico retained a distinct cultural identity, deeply rooted in its Hispanic heritage, language, and traditions.

The economy of Puerto Rico in 1910 was predominantly agrarian, with sugar, coffee, and tobacco serving as the mainstay of its economic output. The presence of the United States led to an increase in sugar production and the modernization of the agricultural sector, which, while boosting the economy, also fostered socioeconomic disparities and led to labor unrest.

Education underwent significant reforms under the American administration, with an emphasis on public schooling and the introduction of English as a medium of instruction. These changes aimed to improve literacy rates and integrate the island into the American way of life, but also sparked debates about cultural assimilation and national identity.

Socially, the early 20th century was a period of vibrant cultural expression for Puerto Rico. Music, literature, and arts flourished, reflecting a society in the midst of transformation, yet steadfast in its cultural expressions. The island's strategic location in the Caribbean also made it a crucial military outpost for the United States, further intertwining its fate with American geopolitical interests.

\section{The Cave}

The cave is located within the municipality of a small town in southern Puerto Rico. It has a significant karst formation, embodying both the geological complexity and the anthropological richness of the region. This cave is part of the extensive limestone karst systems prevalent in Puerto Rico, characterized by their formation through the dissolution of soluble limestone rocks that lead to features such as sinkholes, underground streams and caverns \citep{lace2013coastal}.

The speleological characteristics of the cave, including stalactites, stalagmites, and other speleothems, provide valuable data for the study of geomorphological processes and paleoclimate interpretations. These calcite formations are the result of the deposition of calcium carbonate and other minerals, providing information on the past environmental conditions and hydrogeological dynamics of the area.

Archaeologically, the cave is a site of paramount importance due to its collection of Taíno pictographs and petroglyphs, which contribute to our understanding of the Pre-Columbian cultures of the Caribbean. These petroglyphs, etched into the walls of the caves, serve as a tangible link to the indigenous populations that once thrived in Puerto Rico \citep{schwantes2011caves}. The motifs and symbols depicted offer evidence for religious, social, and possibly astronomical practices of the Taíno people, providing a focal point for ethnohistorical studies \citep{robiou1984astronomy}.

Radiocarbon dating techniques of organic pigment sampled from 11 different rupestrian rock art located in the cave date back to 1230 to 1380 AD \citep{ramos2021aproximacion,acostaabsolute}. The petroglyphs found in the cave are associated with the Taíno and fall in the period of political and regulation nucleation of 1000 to 1500 AD of Borik\'en (Puerto Rico) \citep{rodriguez2014}. 

In the context of the early 20th century, the cave would have been situated in a predominantly agrarian landscape, with the surrounding area undergoing transformations due to shifts in agricultural practices and land use, including cave minning for guano (bat droppings). The cave itself, due to its inaccessibility and relative isolation, remained largely untouched by these changes, preserving its natural and cultural heritage.

From a conservation perspective, the cave is indicative of the broader challenges facing karst regions in Puerto Rico, including habitat destruction, biodiversity loss, and the need for sustainable management practices. The cave ecosystem supports a variety of endemic and specialized species, highlighting the ecological value of subterranean habitats.

Scientifically, this cave offers a multidisciplinary research platform that includes geology, archaeology, biology, and environmental science. Its preservation and study are essential to advance our knowledge of karst landscapes, understand human-environment interactions in the Caribbean, and protect Puerto Rico's natural and cultural heritage for future generations. Based on the importance of this site, it was recommended that the cave be added to the National Register of Historical Sites for its conservation, preservation and protection to the \textit{Oficina Estatal de Conservación Historica of Puerto Rico} \citep{rodriguez2017}.

\section{Methodology}

The research was structured around five key steps: deciphering the cave art, identifying the depicted comet, determining the artist, and uncovering the motivation behind the artwork. Primary data was collected through field investigations, while secondary data was sourced from historical records, including newspapers, censuses, and astronomical reports.

\subsection{Deciphering the Cave Art}

\begin{figure}
\centering
\includegraphics[width=\linewidth]{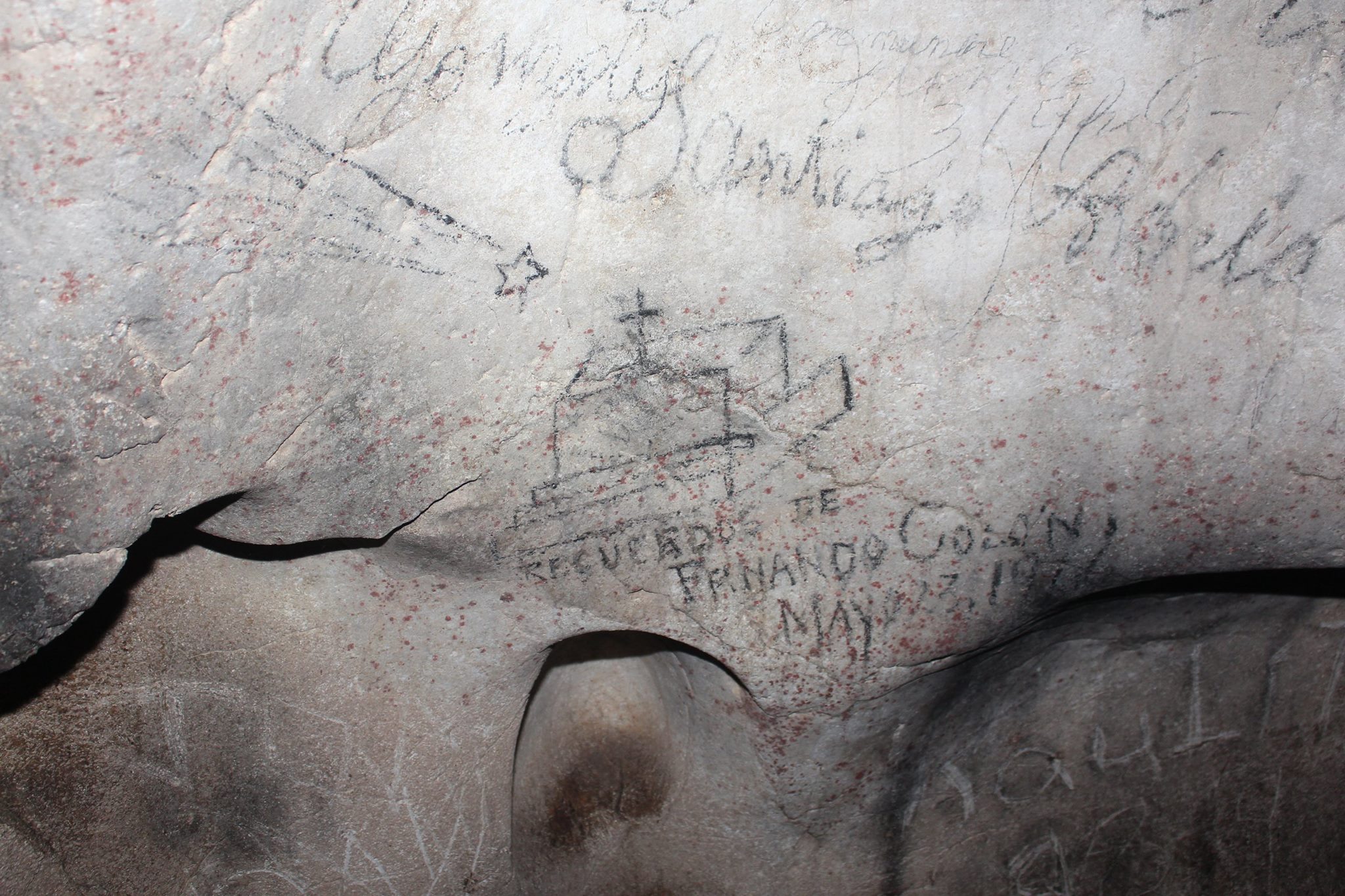}
\caption{Portrait of a cave art depicting a comet over a tomb. Other postdating inscriptions are also visible. \credit\label{fig:art1}}
\end{figure}

An exploration visit to the cave in 2018 revealed an artwork probably made with charcoal. It illustrates a comet or meteor above a structure similar to a church or tomb, signed ``RECUERDOS DE FERNANDO COLÓN, MAY 23, 1910" (Figure \ref{fig:art1}). This suggests a deliberate act of memorialization by the artist, highlighting a lasting impact and significance. The date of inscription on the artwork could not necessarily represent the actual time of creation but rather serve as a commemorative marker, reflecting on the event from a future perspective. We started an investigation on the origin of art in October 2023.

\begin{figure}
\centering
\includegraphics[width=\linewidth]{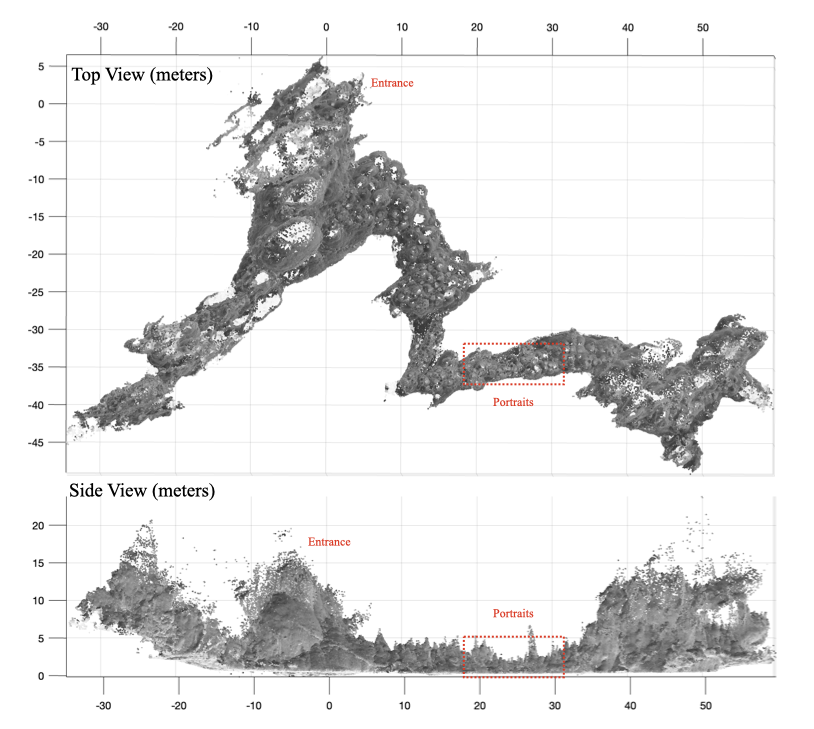}
\caption{Three dimensional lidar top and side renderings of the cave showing the cave entrance and the the narrow passage to the location of the portraits (red dotted box). Scale units are in meters. \credit\label{fig:map}}
\end{figure}

Reaching this area within the cave requires crawling, as it is devoid of natural illumination (Figure \ref{fig:map}). Consequently, the artist would have needed to rely on sources of artificial light, such as lanterns or a makeshift fire, to execute his artwork. The use of artificial lighting in such a secluded and dark environment would have added complexity to the creation of the artwork, highlighting the artist's commitment to commemorating the celestial event despite the physical constraints of the cave's interior.

\begin{figure*}
\centering
\includegraphics[width=\linewidth]{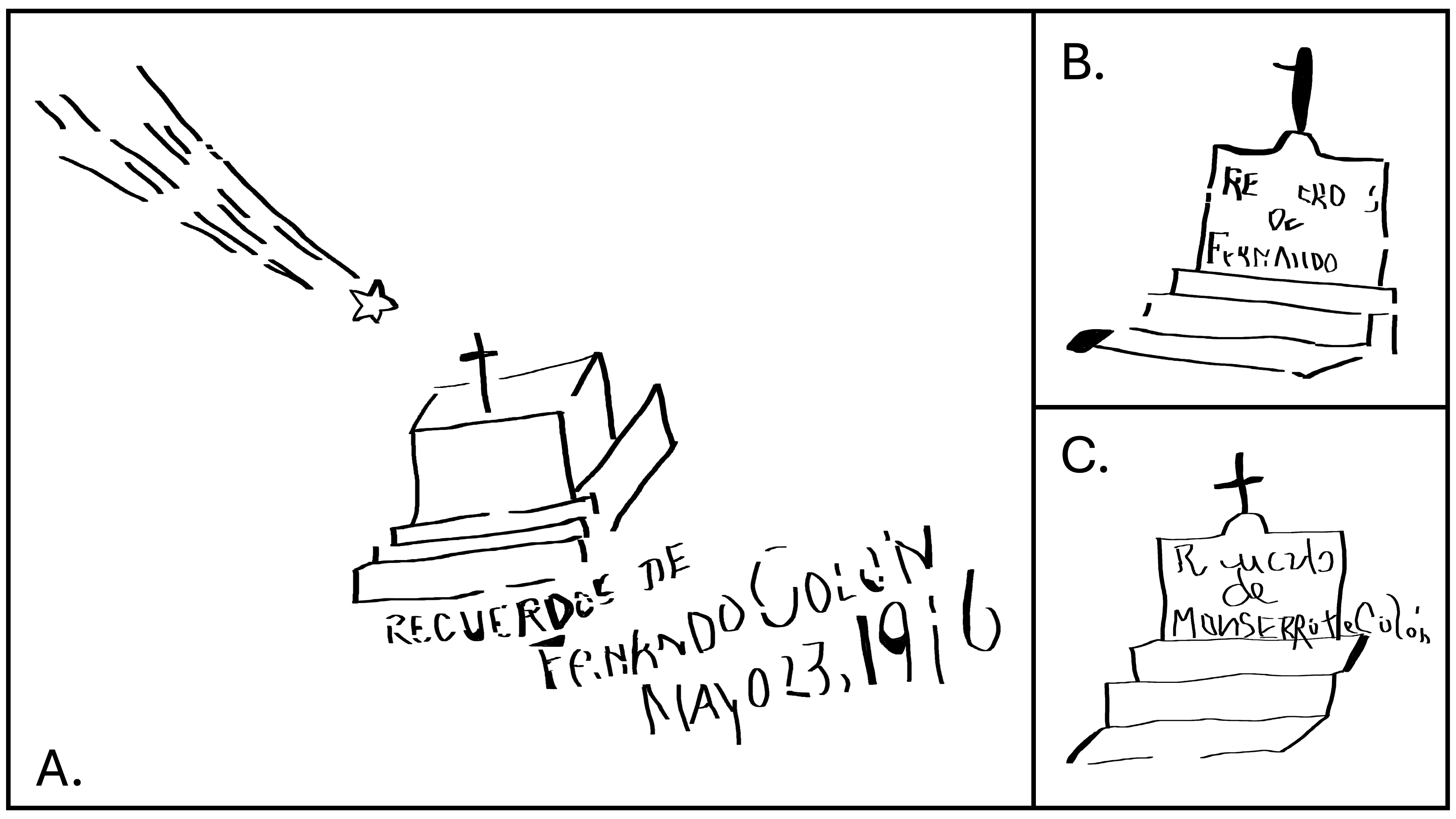}
\caption{Traze of the three cave arts shown in Figures \ref{fig:art1} (A), \ref{fig:art2} (B), and \ref{fig:art3} (C). Images are aproximatelly to scale. \credit\label{fig:traze}}
\end{figure*}

The isometric portrayal of the church shows considerable artistic talent, portraying a structure that more closely resembles a chapel equipped with stairs, or more likely a tomb, pantheon, or mausoleum, rather than a conventional church (Figure \ref{fig:traze}). This suggests not only a personalized artistic interpretation but also an exploration beyond typical ecclesiastical imagery, possibly reflecting a deeper symbolic or personal significance for the artist.

The artist might have drawn inspiration from contemporary illustrations for his depiction of a comet. The iconography of a comet, depicted as a star with a tail, has long been established in artistic representations, suggesting a connection to traditional visual language. Despite an exhaustive search, no direct parallels in existing literature were found, especially not the unique combination of a comet positioned above a religious structure or burial site. Furthermore, the specific church illustrated does not closely match the architectural details of the local town church, which was established in the late 18th century.

The art literacy suggests that the artist was a person of considerable education, given him access to and engagement with newspapers of the time. His meticulous attention to orthographic details, such as accents and commas, further indicates a high level of scholastic achievement. The use of the American date format, specifying month, day, and year, implies familiarity with conventions outside the local norm, possibly reflecting broader educational or cultural exposure. 

This is particularly noteworthy considering that in 1910, only 35.5\% of Puerto Rico's population over the age of ten were literate. The artist's proficiency in reading and writing, therefore, not only sets him apart within his contemporary societal context, but also hints at his privileged access to education and information, marking him as a distinct figure in the historical landscape of early twentieth-century Puerto Rico.

\begin{figure}
\centering
\includegraphics[width=\linewidth]{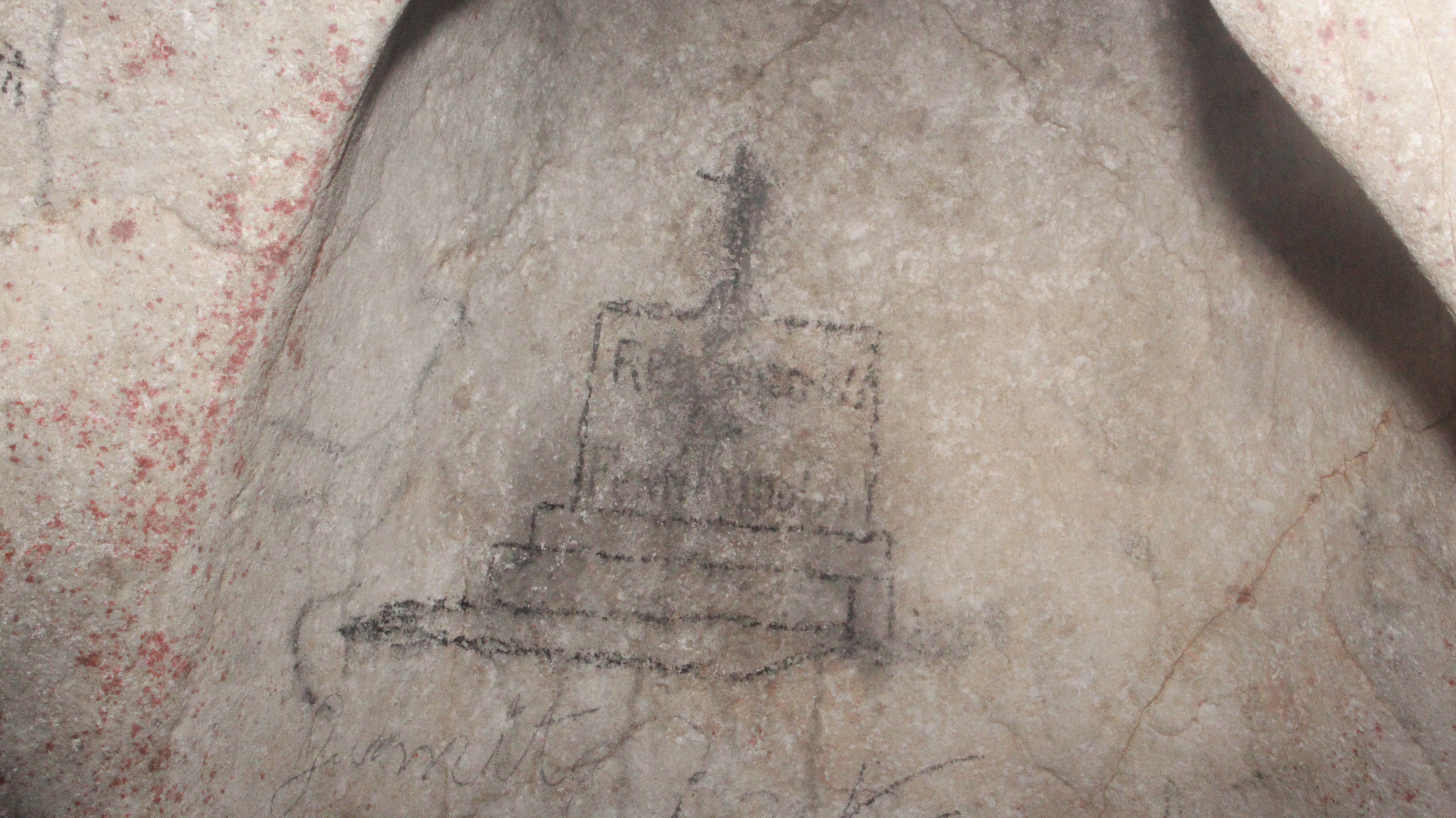}
\caption{A second portrait of a tomb with the inscription "Recuerdos de Fernando". \credit\label{fig:art2}}
\end{figure}

\begin{figure}
\centering
\includegraphics[width=\linewidth]{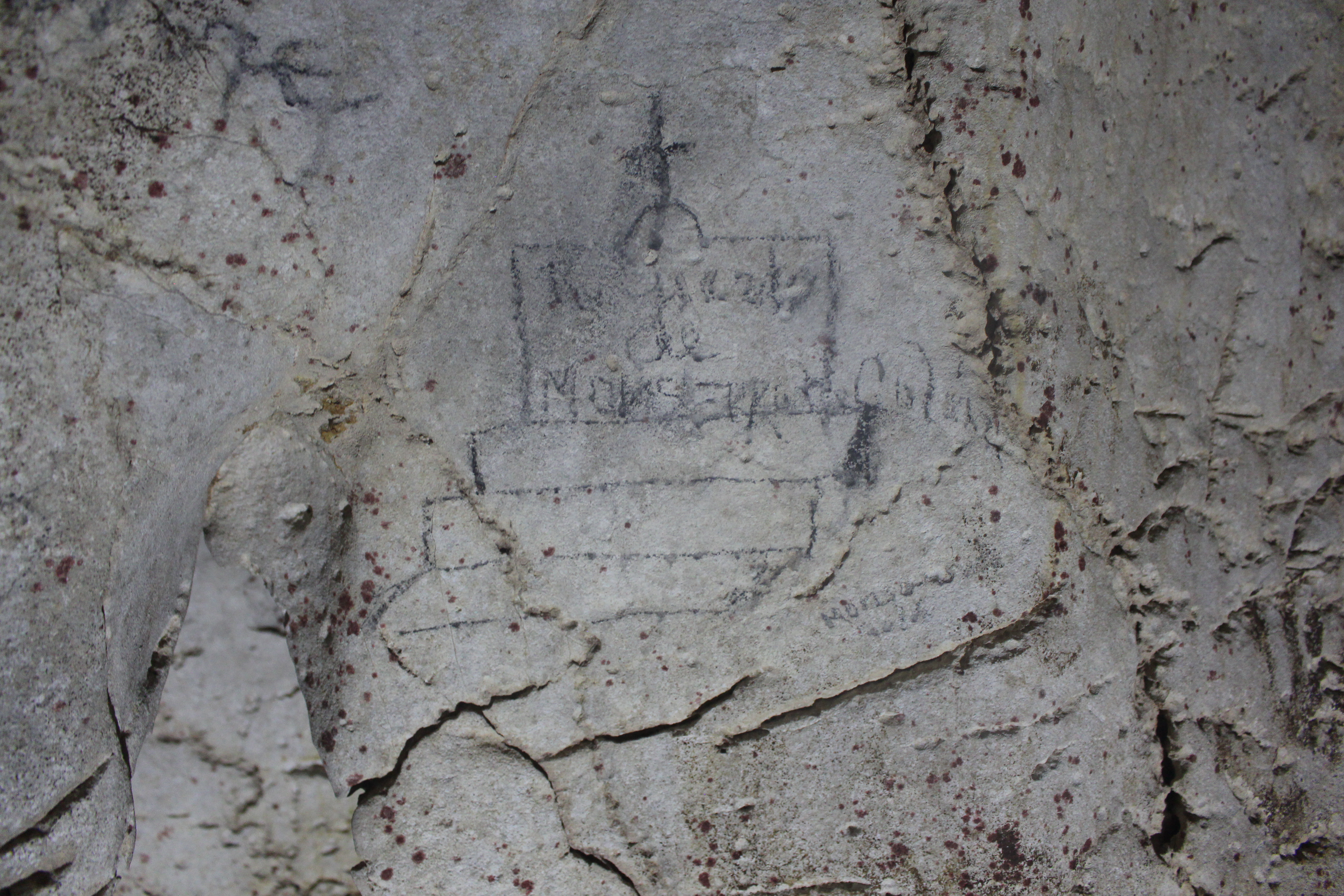}
\caption{A third portrait of a tomb with the inscription "Recuerdo de Monserate Colón". \credit\label{fig:art3}}
\end{figure}

\subsection{Identification of the Comet}

Halley's Comet (1910) is perhaps the most celebrated comet; it made a well-documented pass by Earth in April and May of 1910. Its predictability and the spectacle it provides every 76 years make it a cornerstone of comet observation. The timing and imagery in the cave art align with the passage of Halley's Comet and the associated public fear, substantiated by contemporary media reports. Furthermore, May 1910 aligns with Earth's transit through Halley's Comet's tail.

No other comet visible to the naked eye from May 1910 through 2000 matches the art date as precisely as this occurrence. Comets visible to the naked eye are indeed rare astronomical phenomena, making their appearances significant events for both the scientific community and the general public. The period between 1910 and 2000 saw several such comets, each with its unique characteristics and impact on observers.

Comet Skjellerup-Maristany was discovered in 1927, this comet was visible without the aid of telescopes in May of the same year. Though not as famous as others, its appearance contributed valuable data to the study of comets. Comet Arend-Roland is known for its brilliant display in the night sky of April and May 1957, it featured a luminous nucleus and a pronounced tail, captivating observers worldwide.

Comet Mrkos was discovered in 1957. Emerging in the same year as Comet Arend-Roland, Comet Mrkos graced the skies around May, offering another celestial spectacle for that year. Comet Hale-Bopp (1997) stands out as one of the most widely observed comets of the 20th century, visible to the naked eye for an unprecedented 18 months. Though its closest approach to Earth occurred in March 1997, it continued to be a fixture in the night sky well into the latter part of the year. Comet Hyakutake, with its closest approach in March 1996 was distinguished by its extensive tail and striking blue-green hue.

These comets not only provided spectacular views but also important opportunities for scientific observation and public engagement with astronomy. Their appearances have been etched into the collective memory of those who witnessed them, serving as vivid reminders of the dynamic and ever-changing nature of our universe.

\subsection{The Artist}

\begin{figure*}
\centering
\includegraphics[width=\linewidth]{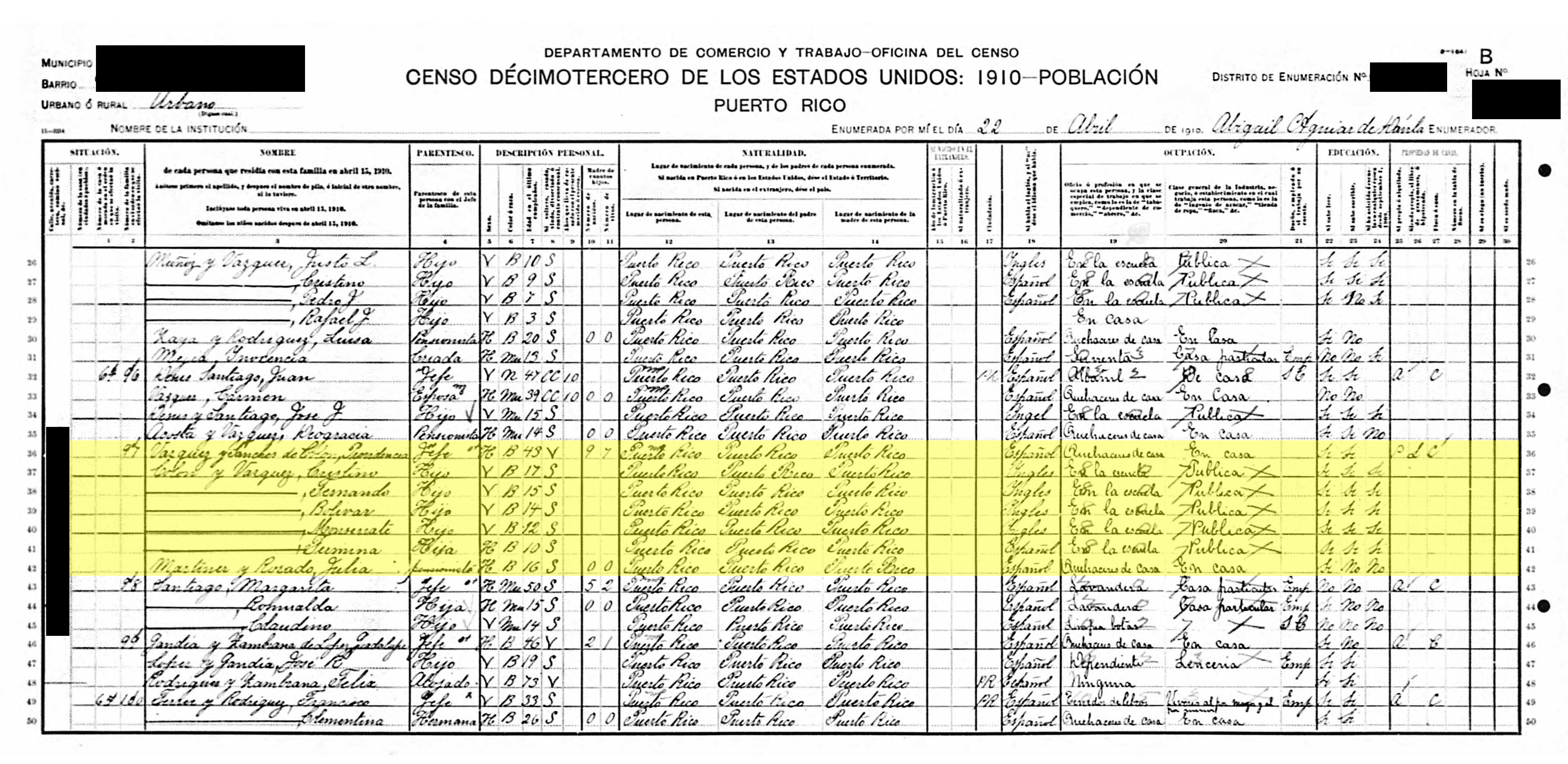}
\caption{1910 Census register with the members of the Colón-Vazquez household (yellow highlight) and neighbors. The location was redacted. \label{fig:census}}
\end{figure*}

The census data examination for the year 1910 showed a mere four individuals bearing the name Fernando Colón within the cave's municipality, with no records of anyone with the same name in the adjacent towns. Among these, a singular Fernando Colón Vázquez distinguished himself not only by his literacy to read and write in both Spanish and English (according to the Census), but also by his residence in a more urbanized setting (Figure \ref{fig:census}). 

This urban dwelling likely gave Fernando improved access to newspapers and periodicals, connecting him more closely to contemporary events and intellectual currents than his namesakes. This detail is crucial, as it not only narrows down the identity of the potential artist behind the cave illustration but also highlights the importance of literacy and media access in early twentieth century Puerto Rican society, especially in terms of engaging with global events like Halley's Comet's passage. A second visit to the cave in November 2023 revealed similar arts with the names Fernando and Monserate Colón (one of his brothers), confirming the connection between the Fernando of the cave and the 1910 census (Figures \ref{fig:art2} and \ref{fig:art3}). 

Fernando Colón Vázquez was a 15-year-old resident of the municipality where the cave was located, living on the main street of the town. His bilingual literacy in Spanish and English distinguished him in a time when such skills were not common, indicating a level of educational attainment that was likely superior within his community. This was an exceptional trait, shared with his siblings, suggesting a family environment that valued education highly. 

Fernando's family background, with his mother managing a household of five children alone following the death of her husband, paints a picture of resilience and dedication to education amidst the challenges of early 20th-century life in Puerto Rico. This context not only highlights the personal achievements of Fernando and his family, but also provides insight into the socio-economic conditions of the time, where widowhood and large families often faced significant hardships.

Throughout his career, Colón engaged in specialized work that required a high degree of literacy and precision, including transcription and typographical roles across multiple organizations. His contributions extended beyond the civilian sector to military service, highlighting his role as a veteran, which would have added a diverse range of experiences to his professional and personal life.

In the latter stages of his career, Colón was employed by the State Insurance Fund, indicating his involvement in the administrative and financial sectors. This detail not only reflects on his professional versatility but also on the socio-economic landscape of Puerto Rico during his lifetime, where veterans often transitioned into pivotal roles within the burgeoning public and private sectors.

Fernando Colón's life spanned from approximately 1896, with his birth date remaining somewhat ambiguous, to his passing in the mid-20th century 1950, marking a life of 54 years. Colón's life and work thus encapsulate a period of significant change and development in Puerto Rican society, with his contributions leaving an indelible mark on the island's historical and cultural fabric.

\subsection{Societal Context and Public Reaction} 

By 1910, Puerto Rico had not yet entered the radio era, which would only commence in 1922 with the inauguration of WKAQ, the island's first radio station. However, the advent of telecommunication through telephones had occurred a decade earlier, around 1900, introducing a modern means of communication. These early telephones, identified by their three-digit numbers, were predominantly accessible in urban centers, reflecting the technological disparities between urban and rural areas at the time.

This urban-centric distribution of telephones underscores the technological and infrastructural development of Puerto Rico in the early 20th century, highlighting a period of gradual modernization limited by geographical and socio-economic factors. The absence of radio broadcasts until the 1920s and the restricted availability of telephones paint a picture of a society on the cusp of a communication revolution, with significant implications for information dissemination and social connectivity across the island.

In 1910, Puerto Rican media extensively covered Halley's Comet's approach, sparking discussions on its potential catastrophic impact on Earth. Notably, \textit{La Correspondencia de Puerto Rico} highlighted concerns over the comet's lethal consequences in an early January edition. Meanwhile, \textit{Puerto Rico Ilustrado} adopted a more lighthearted tone by March, jestingly suggesting the uncertainty of survival post-May, directly alluding to the comet's expected passage.

These contrasting approaches in reporting underscore the wide spectrum of public sentiment and reaction to the comet, from genuine anxiety over its perceived threats to more satirical interpretations of the end-of-times fears. This coverage reflects the cultural and social milieu of early 20th-century Puerto Rico, where the intersection of science, sensationalism, and humor shaped the collective experience of astronomical phenomena.

\section{Discussion}

The artwork's style, lack of similar contemporary images, and the unique depiction of the comet over a non-specific tomb-like structure suggest a personal interpretation of the event by Colón, rather than a direct replication of existing illustrations.

Spanning roughly 9 kilometers along the primary routes, the journey from the home of Fernando Colón Vázquez to the cave represents a significant trek, requiring about two hours on foot. The popular means of transportation in those times were bicycles and horses, although the first cars were introduced five years earlier in 1905. This distance reflects not only the physical separation between his residential area and the cave but also underscores the effort and determination involved in reaching the cave.

The decision to undertake such a journey, in difficult karst topography and semi-arid climate, where modern transportation options were limited, highlights the importance of the destination or the motivation behind the visit. Given the historical context and the effort required to access the cave, this journey would have been a considerable undertaking, suggesting a strong incentive or a significant purpose behind Fernando Colón Vázquez's visit to the cave.

The widespread apprehension regarding the toxic gases associated with Halley's Comet's tail might have driven Fernando Colón, potentially accompanied by his family, neighbors, or many others to undertake a two-hour journey to the deep dark zone of the cave as a refuge. This action, taken in May, suggests a significant level of concern over the celestial event's potential effects.

The determination to isolate themselves in the cave, potentially for an extended period, indicates a profound impact of the comet's passing on the collective psyche of the time, revealing how celestial phenomena could influence human behavior. This episode underscores the interplay between scientific understanding, or lack thereof, and societal reactions to astronomical events, highlighting the lengths to which individuals would go to protect themselves from perceived cosmic threats.

\section{Conclusion}

Fernando Colón Vázquez, a 15-year-old literate resident of southern Puerto Rico, was identified as the artist behind Halley's Comet over a tomb. His education level and access to the contemporary media are highlighted as critical factors enabling his awareness and reaction to the visit of the comet in 1910. The widespread panic over the comet, fueled by sensational media reports predicting catastrophic outcomes, probably motivated Colón Vazquez, his family, and possibly others to seek refuge in the cave.

The case of Fernando Colón's cave art illustrates the intersection of astronomical phenomena, societal fear, and individual expression. This incident sheds light on the broader impacts of celestial events on human societies, particularly in isolated communities with limited scientific understanding. His descendants might provide further details of this event.

The 1986 return of the comet was received with more enthusiasm worldwide \citep{seargent2008halley}. The next visit will be in summer 2061. The 1910 Halley's Comet sighting in Puerto Rico, as captured by Fernando Colón in the cave, is another example of the blend of fear, fascination, and artistic expression that such events can provoke.

\begin{acks}
This research was supported by The Planetary Habitability Laboratory and the Center for Research and Creation of the University of Puerto Rico at Arecibo. In addition, Acosta-Col\'on would like to thank Dariliz Montero, Heriberto Acevedo and Javier Rivera, students of Dr. Reniel Rodr\'iguez Ramos's class (ANTR3025, Principles of Arqueology) at University of Puerto Rio at Utuado, for their help in this project.
\end{acks}

\theendnotes

\bibliography{halley}{}
\bibliographystyle{SageH}

\end{document}